\documentclass[runningheads]{llncs}
\usepackage[T1]{fontenc}
\usepackage{graphicx}
\usepackage{amsmath,url,amssymb} 
\usepackage{cite}
\usepackage{microtype}
\usepackage{booktabs}
\makeatletter
\def\blfootnote{\xdef\@thefnmark{}\@footnotetext}
\makeatother
\usepackage{fancyhdr}
\providecommand{\runtitle}{Undefined}
\providecommand{\runauthor}{Undefined}
\renewcommand{\runtitle}{Filling MIDI Velocity}
\renewcommand{\runauthor}{Z. He et al.}
\begin{document}
\title{Filling MIDI Velocity using U-Net Image Colorizer}
%

%
\author{Zhanhong He\inst{1,2}\orcidID{0000-0002-8940-8437} \and
David Cooper\inst{2}\orcidID{0009-0008-9805-8943} \and
Defeng Huang\inst{1}\orcidID{0000-0002-1431-8859} \and Roberto Togneri\inst{1}\orcidID{0000-0002-3778-4633}
}
\institute{University of Western Australia, Perth WA 6000, Australia \and
Dolby Laboratories, Sydney NSW 2000,  Australia\\
\email{zhanh.he.uwa@gmail.com, david.cooper@dolby.com, \{david.huang,\,roberto.togneri\}@uwa.edu.au}
}
%
\maketitle              
%
\blfootnote{\includegraphics[scale=0.25]{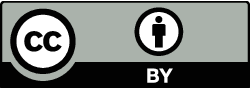} All rights remain with the authors under the Creative Commons Attribution 4.0 International License (CC BY 4.0).\\
Proc. of the 17th Int. Symposium on Computer Music Multidisciplinary Research,\\
London, United Kingdom, 2025}
\vspace{-18pt}
\begin{abstract}
Modern music producers commonly use MIDI (Musical Instrument Digital Interface) to store their musical compositions. However, MIDI files created with digital software may lack the expressive characteristics of human performances, essentially leaving the velocity parameter—a control for note loudness—undefined, which defaults to a flat value. The task of filling MIDI velocity is termed MIDI velocity prediction, which uses regression models to enhance music expressiveness by adjusting only this parameter. In this paper, we introduce the U-Net, a widely adopted architecture in image colorization, to this task. By conceptualizing MIDI data as images, we adopt window attention and develop a custom loss function to address the sparsity of MIDI-converted images. Current dataset availability restricts our experiments to piano data. Evaluated on the MAESTRO v3 and SMD datasets, our proposed method for filling MIDI velocity outperforms previous approaches in both quantitative metrics and qualitative listening tests.

\keywords{MIDI velocity prediction \and U-Net \and Image colorization \and Music expressiveness.}
\end{abstract}
\section{Introduction}\label{sec:introduction}

MIDI (Musical Instrument Digital Interface), acting as digital sheet music playable by machines and software, is the dominant format in modern music production. A MIDI file resembling sheet music sounds mechanical due to the undefined velocity parameter, which defaults to a flat value. In contrast, as shown in Figure 1, MIDI files recorded from human performances capture performer skills through subtle timing and loudness variations, which infuse expressiveness \cite{eduardo2018computational}. Today, music producers are not always masterful in playing musical instruments \cite{hracs2012creative, hong2018but}, and low cost MIDI keyboards may lack sophisticated touch-sensitive sensors. This leads to a demand for automated systems designed to enhance the expressiveness of MIDI compositions.

Enhancing the expressiveness of existing MIDI files is one important focus in music generation \cite{oore2020time}. However, many of these systems modify multiple aspects of MIDI simultaneously \cite{brunner2018midivae, jeong2019virtuosonet, rhyu2022sketching, tang2023reconstructing, borovik2023scoreperformer, zhang2024dexter}, including note timing and loudness, and sometimes note quantity, which can introduce unwanted alterations. The loudness of each music note in a MIDI file is governed by a parameter called MIDI velocity. Studies \cite{simoes2019deep} have shown that rendering only the MIDI velocity enhances expressiveness while preserving the original timing, making it a precise and controllable solution. This task of filling or rendering MIDI velocity has been treated as a sequential prediction problem by previous studies, which employed autoencoders \cite{kuo2021velocity} and sequential models \cite{kim2023piano}.


\begin{figure}[h]
\begin{center}
\includegraphics[width=0.92\textwidth]{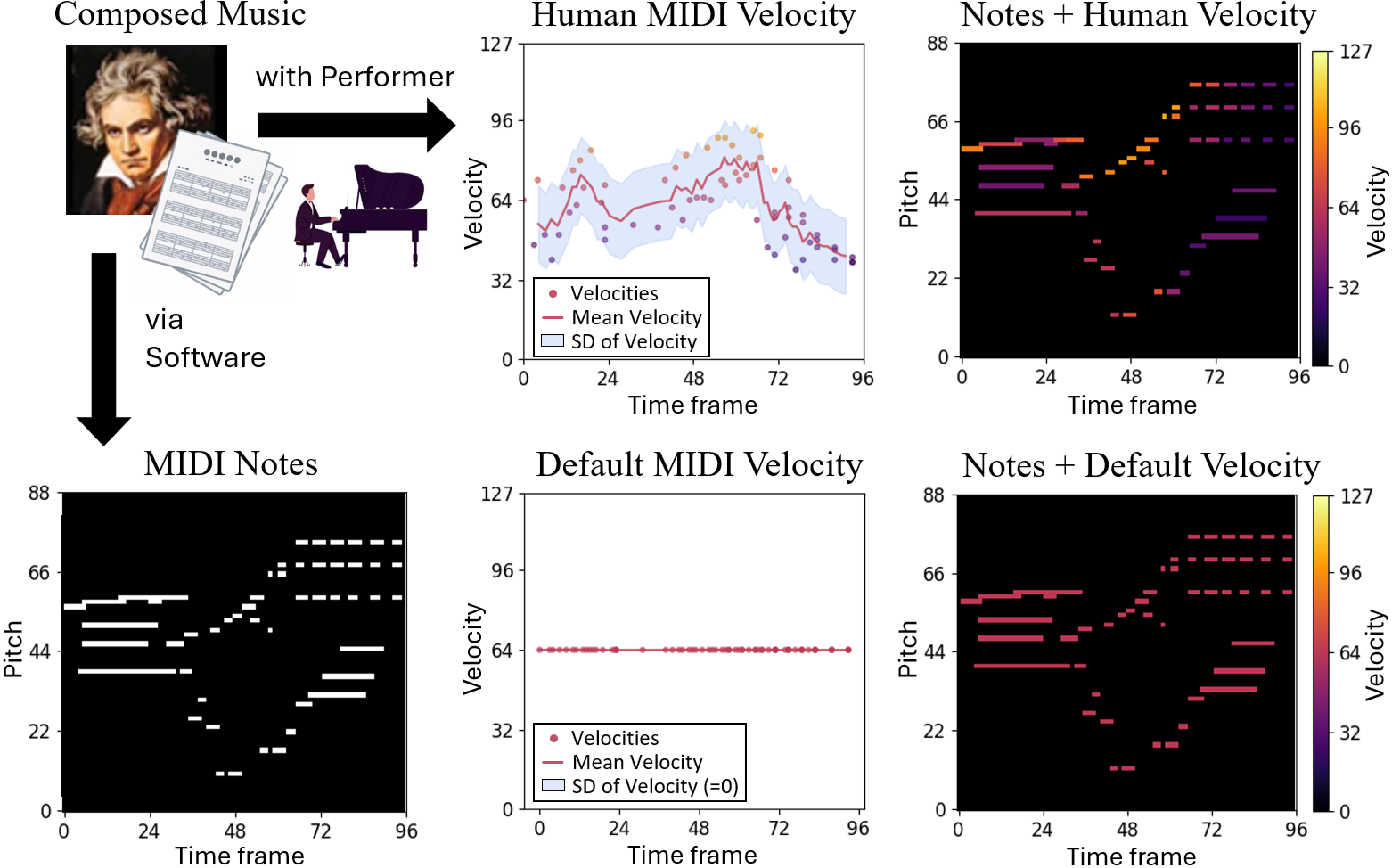} 
\caption{Comparison between MIDI notes with human performed velocity versus Music Software default velocity (64 if user not specified). The standard deviation of velocity (\( SD_{\text{velo}}\)) represents the dispersion of velocities around their mean across the pitches.}
\label{fig:intro}
\end{center}
\end{figure}

Inspired by image colorization \cite{anwar2024imagesurvey}, we reframe MIDI velocity prediction as an image colorization problem, representing MIDI without velocity as a binary pianoroll and target velocity as a colored pianoroll. Image-based methods suit this case well, as they effectively capture the polyphonic structure of instruments such as the piano and guitar, which produce multiple simultaneous notes. While our work focuses on piano data, where well-annotated velocity datasets are most common, the universal nature of MIDI velocity makes cross-instrument generalization a promising direction for future research \cite{gaps2024}.

In this paper, we introduce the U-Net architecture to MIDI velocity prediction, leveraging its success in image colorization, and incorporate window attention to handle the sparsity of MIDI data. In addition, we design a custom loss function tailored to our task. The resulting model is evaluated through both objective quantitative metrics and qualitative assessments via a subjective listening test.

\section{Related Works}\label{sec:related}


How to render a score to be more expressive (i.e., a performance-like MIDI) has been a long-standing topic in music research \cite{oore2020time}. A central goal of this research, MIDI velocity prediction, has been to independently modify MIDI velocity. Early efforts to this problem involved linear basis models \cite{grachten2012linear} and restricted Boltzmann machines \cite{vanherwaarden2014predicting}. More recently, Kuo et al. \cite{kuo2021velocity} implemented a convolutional autoencoder (ConvAE), while a Seq2Seq model \cite{kim2023piano} reported the best results by integrating Luong attention into a BiLSTM.

While recent methods have treated MIDI as a sequence, those sequential models prioritize global features over local details \cite{li2024cnntransformer}, potentially affecting the scattered distribution of velocities. Inspired by image colorization, where precise grayscale images are overlaid with blurred color predictions \cite{zhang2016colorful}, MIDI velocity prediction can be approached similarly by leveraging given MIDI notes. This suggests that U-Net, a widely used architecture in image colorization \cite{wang2022cunet}, can be effective for MIDI velocity prediction. U-Net also dominates image segmentation and is frequently combined with self-attention mechanisms \cite{siddique2021unetvar, azad2024unetreview}.

Both U-Net and attention mechanisms have shown success in music information retrieval (MIR) research. While U-Net has been effective in automatic music transcription (AMT) \cite{ped2020presunet, scarp2023uetamt}, the attention mechanism has been used to refine velocity estimation from performance audio \cite{kim2024method}. Since our task is MIDI-only, their audio-dependent approach is not applicable.

\section{Methods}\label{sec:method}

\subsection{Matrix Representation}
To process MIDI as images, we convert the MIDI into a three matrices with \(T \times P\) dimension, where \( T \) is the number of time frames and \( P=88 \) is the number of pitches. The three matrices include a binary onset roll \( O \) marking note starts, a binary frame roll \( F \) indicating note activation over time, and a velocity roll \( V \) acting as color intensity. For the velocity roll, integer values [0,127] are normalized to the range [0,1) to align with the model's output activation layer. The final integer velocities will denormalize by scaling and rounding the model's output. As shown in Figure 2, these matrices are highly sparse.

\subsection{MIDI Segmentation}\label{subsec:midi_seg}
The MIDI segment duration is a key consideration in our approach. Since a MIDI file often exceeds 3 minutes in length, we split it into short segments to manage computational load. The number of time frames \( T \) defines the temporal granularity of the MIDI-converted matrix, so the timestep resolution is given by:
\begin{equation}
\text{Resolution} = \frac{\text{Segment\ Duration}}{T}
\label{eq:timing_resolution}
\end{equation}
Unlike tasks that require high timestep resolution for precise event detection, our task leverages known MIDI timings, making a computationally efficient timestep resolution feasible. With a fix input size \( T=96 \) to keep size affordable, we experimented the segment durations of {5s, 10s, 15s, 20s} and found that 10 seconds yielded the best results, as detailed in our hyperparameter search.\footnote{wandb report has concluded our experiment history of hyperparameter searching, available at: \url{https://api.wandb.ai/links/zhanh-uwa/wpzvcb76}} A 10-second segment likely provides superior semantic context by encapsulating a complete musical phrase (e.g., four measures at 120 BPM) compared to other durations we tested.


\begin{figure}[h]
\begin{center}
\includegraphics[width=\textwidth]{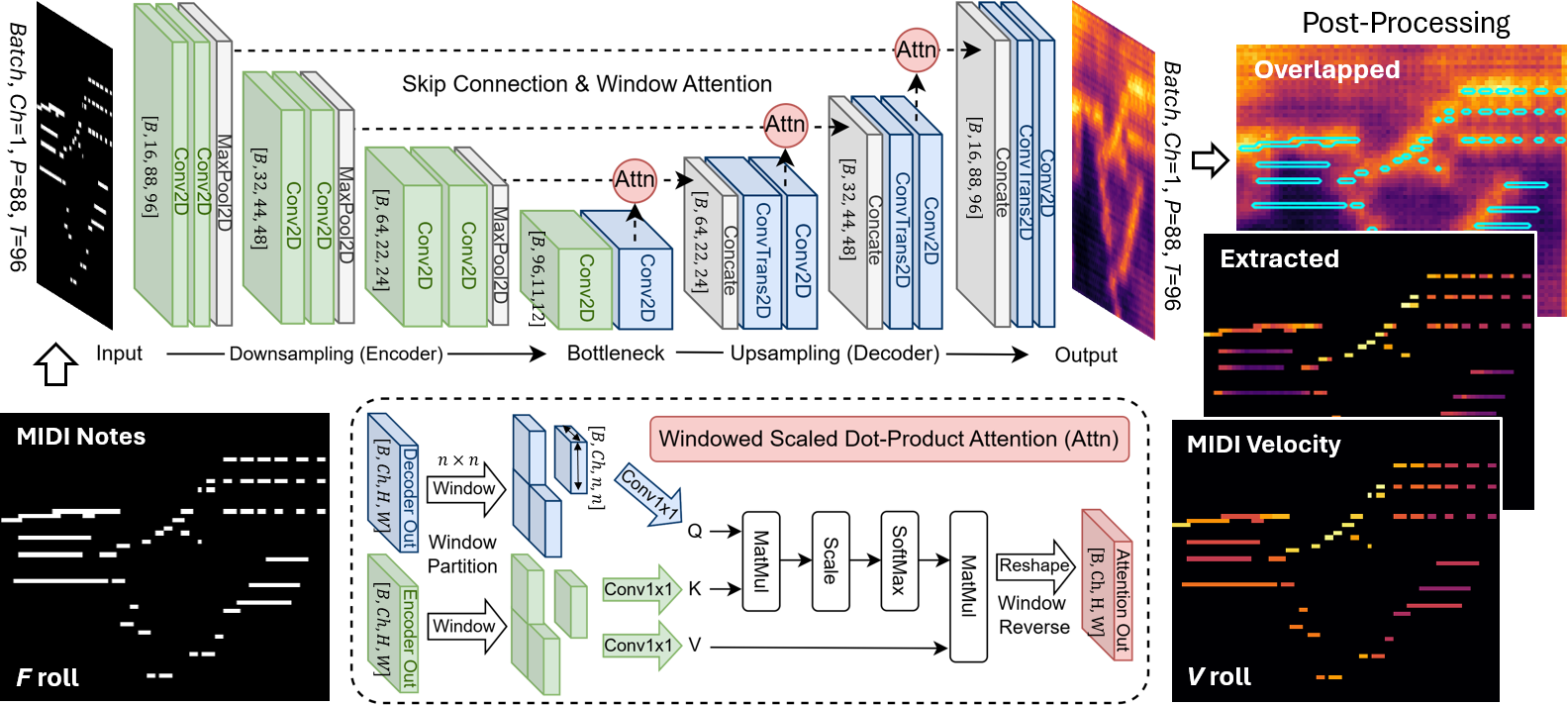}
\caption{Proposed U-Net architecture. Model input (\(F\) roll) comprises 88 pitch bins and 96 time frames. Attn block denotes the windowed scaled dot-product attention. The final velocity roll (\(V\) roll) is generated during post-processing by extracting velocity at note positions, and then assigning each note the velocity at its onset.}
\vspace{-24pt}
\label{fg:model}
\end{center}
\end{figure}

\subsection{Model Architecture}

The proposed architecture is shown in Figure \ref{fg:model}. The U-Net extracts higher-level features through downsampling, with skip connections preserving details and global patterns. All convolution blocks use $3 \times 3$ kernels, stride 1, and padding 1, followed by sigmoid activation and batch normalization. To scale features by a factor of 2, we perform downsampling with a standard $2 \times 2$ MaxPool2D layer; for upsampling, we employ a transpose convolution with $4 \times 4$ kernels and stride 2, a distinctive strategy introduced in \cite{wang2022cunet}.

To handle MIDI data sparsity, we integrate the windowed scaled dot-product attention (window attention) \cite{shen2023linear} into our U-Net. Traditional self-attention \cite{vas2017attn} operates on the full feature map $X \in \mathbb{R}^{H \times W}$, where $H$ and $W$ denote the height and width of the attention inputs, as depicted in Figure 2. Window attention partitions $X$ into $n \times n$ non-overlapping windows and computes attention within each. This reduces computational complexity while enhancing feature aggregation \cite{chen2022hts}. 

The window size ($n$) is a tunable hyperparameter balancing local and long-range dependencies. We explored $n \in \{1, 2, 4, 8\}$ and found that a $2 \times 2$ window attention yielded the best performance (see wandb report\footnotemark[3]). 
This suggests that while window attention is effective, larger windows can over-compress information and reduce effectiveness.

\subsection{Loss Function}

The proposed loss function combines binary cross-entropy (BCE) loss with cosine similarity (CosSim) introduced in \cite{kim2023piano}, with \(\alpha=0.2\), defined as:
\begin{equation}
\mathcal{L}_{\mathrm{Combine}} = (1-\alpha) \, \mathcal{L}_{\mathrm{BCE}} + \alpha \, \left(1-\mathrm{CosSim}\right)
\end{equation}
where the BCE loss is used to optimize the prediction error; CosSim is computed for each pitch and then averaged, capturing the trending of velocity changes over time:
\begin{gather}
\text{CosSim} = \frac{1}{P}\sum_{p=1}^{P} 
\frac{\sum_{t=1}^{T} y_{t,p}\,\hat{y}_{t,p}}
{\sqrt{\sum_{t=1}^{T} y_{t,p}^{2}}\;\sqrt{\sum_{t=1}^{T} \hat{y}_{t,p}^{2}}} \\
\mathcal{L}_{\mathrm{BCE}} = \frac{1}{TP}\sum_{t=1}^{T} \sum_{p=1}^{P} 
l_{\text{bce}}\!\left(y_{t,p}, \hat{y}_{t,p}\right)
\end{gather}
here, CosSim and BCE are functions pre-built in PyTorch, with $y_{t,p}$ and $\hat{y}_{t,p}$ denoting the target and predicted velocity, respectively. The indices $t$ and $p$ represent the time and pitch dimensions. To deal with the sparsity, we apply a masking operation <$m$> using the onset roll \(O\), which ignores silent time steps and counts each note once at its onset. In addition, a weighting operation <$w$> is introduced to reduce the boundary velocity prediction errors emphasized in \cite{jeong2020note, kim2024method, kim2023score}. Motivated by the Gaussian distribution of velocity observed in Figure 3, we design a V-shaped weighting centered on 64 (normalized to 0.5), with an empirical factor of 3 to enhance regions away from the midpoint. The updated BCE loss with masking and weighting is defined as: 
\begin{equation}
\mathcal{L}_{\mathrm{BCE}}^{<m,w>} 
= \frac{1}{TP}\sum_{t=1}^{T} \sum_{p=1}^{P} 
O_{t,p} \cdot \bigl(1 + 3 \lvert V_{t,p} - 0.5 \rvert \bigr) \cdot 
l_{\text{bce}}\!\left(y_{t,p}, \hat{y}_{t,p}\right),
\end{equation}
in which $O_{t,p}$ is the onset-roll mask and $(1 + 3|V_{t,p} - 0.5|)$ is a weighting factor based on velocity roll $V$. Finally, $\mathcal{L}_{\mathrm{BCE}}$ in Eqn (2) is replaced with $\mathcal{L}_{\mathrm{BCE}}^{<m,w>}$ to form $\mathcal{L}_{\mathrm{Combine}}^{<m,w>}$. The effectiveness of this loss was validated in our wandb report.\footnotemark[3]




\section{Experiment}\label{sec:experiment}
\subsection{Dataset}

For training, we use the MAESTRO v3.0.0 dataset \cite{maestro}, recorded by skilled pianists on Yamaha Disklavier pianos during the International Piano-e-Competition. The default train/valid/test split is used. With 1,276 performances totaling over 200 hours, MAESTRO provides an ideal foundation for modeling MIDI velocity.

\begin{figure}[t]
\centering
\includegraphics[width=0.96\textwidth]{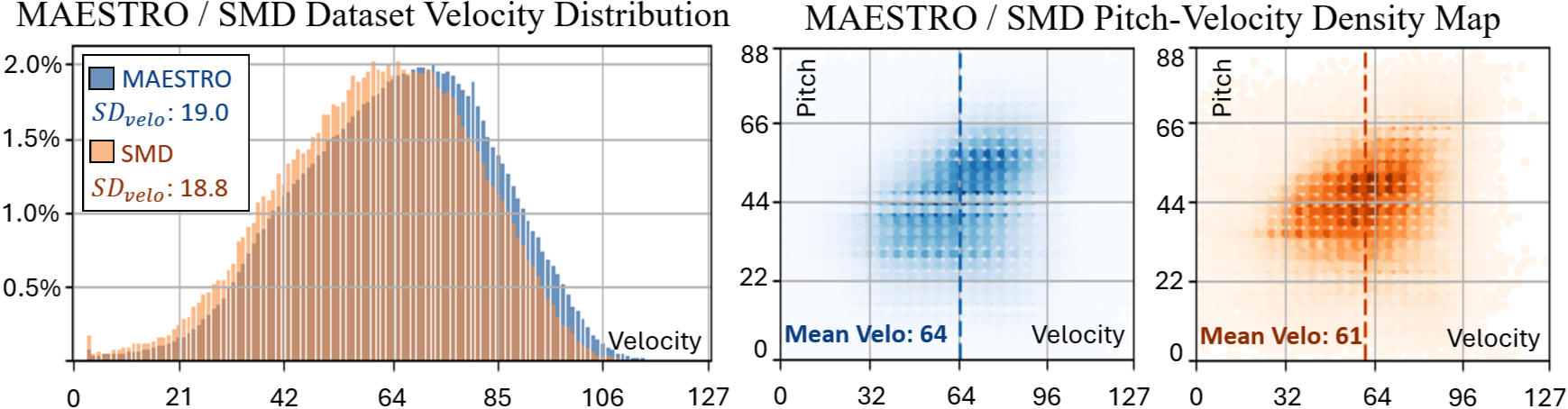}
\caption{MIDI data distributions of the MAESTRO (blue) and SMD (orange) datasets, with density maps highlighting the similarity in their MIDI feature correlations.}
\label{fg:data_dist}
\end{figure}

For evaluation, we use the Saarland Music Data (SMD) dataset \cite{smd}, which comprises 50 performances also recorded on a Yamaha Disklavier. We selected SMD for cross-dataset evaluation to assess model generalization, instead of the Piano-e-Competition dataset used in \cite{kuo2021velocity, kim2023piano} which has significant performance overlap with MAESTRO. The suitability of SMD is confirmed in Figure \ref{fg:data_dist}. Furthermore, SMD is used for qualitative assessment through a subjective listening test of 8 selected performances, as listed in Table 1. As SMD only has composer-style overlaps with MAESTRO (none of performance overlap), it allows us to evaluate our model on both seen and unseen compositional styles.



\vspace{-0.3cm}
\begin{table}[h]
\centering
\small 
\caption{Selected SMD performances for the subjective listening test, with SMD composer statistics and their overlap with the MAESTRO train set.}
\label{tb:composertb}
\begin{tabular}{|l|l|l|l|l|}
\hline
& \multicolumn{2}{c|}{\textbf{MAESTRO train set}} & \multicolumn{2}{c|}{\textbf{SMD dataset}} \\
\hline
\textbf{Composer} & \textbf{Total Perf.} & \textbf{Total Dur.} & \textbf{Total Perf.} & \textbf{Selected Perf.} \\
\hline
Chopin & 145 & 19.9 h & 13 & Op010-04 \\
Bach & 114 & 11.2 h & 8 & BWV849-02 \\
Beethov. & 110 & 20.5 h & 7 & Op027No1-01 \\
Liszt & 93 & 16.0 h & 3 & - \\
Schuman. & 33 & 12.4 h & 3 & - \\ \hline
Rachman. & 29 & 4.2 h & 3 & Op036-02 \\
Haydn & 29 & 3.7 h & 4 & Hob017No4 \\
Mozart & 27 & 3.9 h & 2 & KV265 \\
Scriabin & 22 & 4.1 h & 1 & - \\
Brahms & 20 & 6.1 h & 3 & - \\ \hline
Bartok & 0 & 0 h & 3 & SZ080-03 \\
Ravel & 0 & 0 h & 2 & JeuxDEau \\
\hline
\end{tabular}
\end{table}
\vspace{-18pt}

\subsection{Training Setup}
The models we trained include the proposed U-Net and a re-implemented ConvAE \cite{kuo2021velocity}. Following the training strategy of \cite{zhang2023symbolic}, we arranged continuous segments of a song into the same batch to preserve the musical structure, thereby aiding semantic learning. Both models were trained for 300 epochs on the MAESTRO train set with a learning rate of 1e-5, a batch size of 3, and the same loss function. Training took approximately 12 hours on an NVIDIA V100 32GiB GPU using the Ranger21 optimizer. The top three checkpoints of each model, selected based on the MAESTRO validation set performance, were tested on the MAESTRO test set and the SMD dataset for cross-dataset evaluation.

\subsection{Evaluation Metrics}

All objective evaluation metrics are computed on the denormalized MIDI velocity, restored to the original scale of 0 to 127. We adopt the mean absolute error (MAE), mean square error (MSE), and standard deviation of velocity (\( SD_{\text{velo}}\)), which are standard metrics in MIDI velocity prediction \cite{kuo2021velocity, kim2023piano}. We also incorporate the standard deviation of absolute error (\( SD_{\text{ae}}\)) and Recall, both prevalent in similar research \cite{toyama2023automatic, kim2024method}. The Recall uses a standard 10\% error tolerance.

The subjective listening test follows the mean opinion score (MOS) of MUSHRA \cite{series2014mursha} framework. Participants rated the expressiveness of MIDI-generated audio on a 100-point scale, mapped to values from 0 to 5, with five labeled intervals (from "bad" to "excellent") for ease of use.

\section{Results and Discussion}\label{sec:results}
\subsection{Quantitative Results}

Tables \ref{tb:result_smd} and \ref{tb:result_maestro_test} present the model performance, with all models trained exclusively on the MAESTRO train set. The Flat model assigns a fixed velocity of 64, representing default music software behavior. The Seq2Seq model uses pretrained weights from \cite{kim2023piano}, while ConvAE \cite{kuo2021velocity} is re-implemented and trained in our framework. Both tables demonstrated that the proposed U-Net outperformed other models across all objective metrics.

\begin{table}[h]
\centering
\small
\caption{Quantitative results on the SMD dataset, where \(\uparrow\) and \(\downarrow\) indicate whether higher or lower values are better.}
\begin{tabular}{lccccc}
\toprule
\textbf{Model} & \textbf{MAE \(\downarrow\)} & \textbf{MSE \(\downarrow\)} & \textbf{SD\textsubscript{ae} \(\downarrow\)} & \textbf{Recall \(\uparrow\)} & \textbf{SD\textsubscript{velo} \(\uparrow\)} \\
\midrule
Flat (all velocities set to 64) & 15.3 & 367.5 & 10.4 & 49.2\% & 0 \\
Seq2Seq \cite{kim2023piano} & 15.1 & 356.9 & 10.9 & 48.8\% & 8.5 \\
ConvAE \cite{kuo2021velocity} (re-implemented) & 12.5 & 258.1 & 9.6 & 58.5\% & 9.8 \\
\textbf{U-Net (proposed)} & \textbf{11.2} & \textbf{217.5} & \textbf{9.2} & \textbf{65.1\%} & \textbf{11.1} \\
\bottomrule
\end{tabular}
\label{tb:result_smd}
\end{table}

\begin{table}[h]
\centering
\small
\caption{Quantitative results on the MAESTRO test set, where \(\uparrow\) and \(\downarrow\) indicate whether higher or lower values are better.}
\begin{tabular}{lccccc}
\toprule
\textbf{Model} & \textbf{MAE \(\downarrow\)} & \textbf{MSE \(\downarrow\)} & \textbf{SD\textsubscript{ae} \(\downarrow\)} & \textbf{Recall \(\uparrow\)} & \textbf{SD\textsubscript{velo} \(\uparrow\)} \\
\midrule
Flat (all velocities set to 64) & 14.8 & 333.8 & 10.2 & 48.5\% & 0 \\
Seq2Seq \cite{kim2023piano} & 13.5 & 286.1 & 9.9 & 53.8\% & 6.2 \\
ConvAE \cite{kuo2021velocity} (re-implemented) & 12.3 & 250.7 & 9.7 & 59.4\% & 9.7 \\
\textbf{U-Net (proposed)} & \textbf{11.5} & \textbf{226.2} & \textbf{9.4} & \textbf{63.8\%} & \textbf{10.7} \\
\bottomrule
\end{tabular}
\label{tb:result_maestro_test}
\end{table}

The validity of evaluation metrics warrants further discussion. When visualizing the results, as demonstrated in Figure 4, we found: (1) Accuracy metrics (including MAE, MSE, \(SD_{\text{ae}}\), and Recall) usually show significant differences, but not in certain cases (e.g., the Chopin Op10-04 segment), suggesting an artifact caused by a local optimum when mid-value predictions dominate. (2) \( SD_{\text{velo}} \) is a key differentiator that effectively reflects human-likeness. By capturing the dispersion of MIDI velocity around its mean, higher values reveal a greater clarity between the left and right hands, potentially enhancing expressiveness. 


\vspace{-6pt}
\begin{figure}[h]
\begin{center}
\includegraphics[width=\columnwidth]{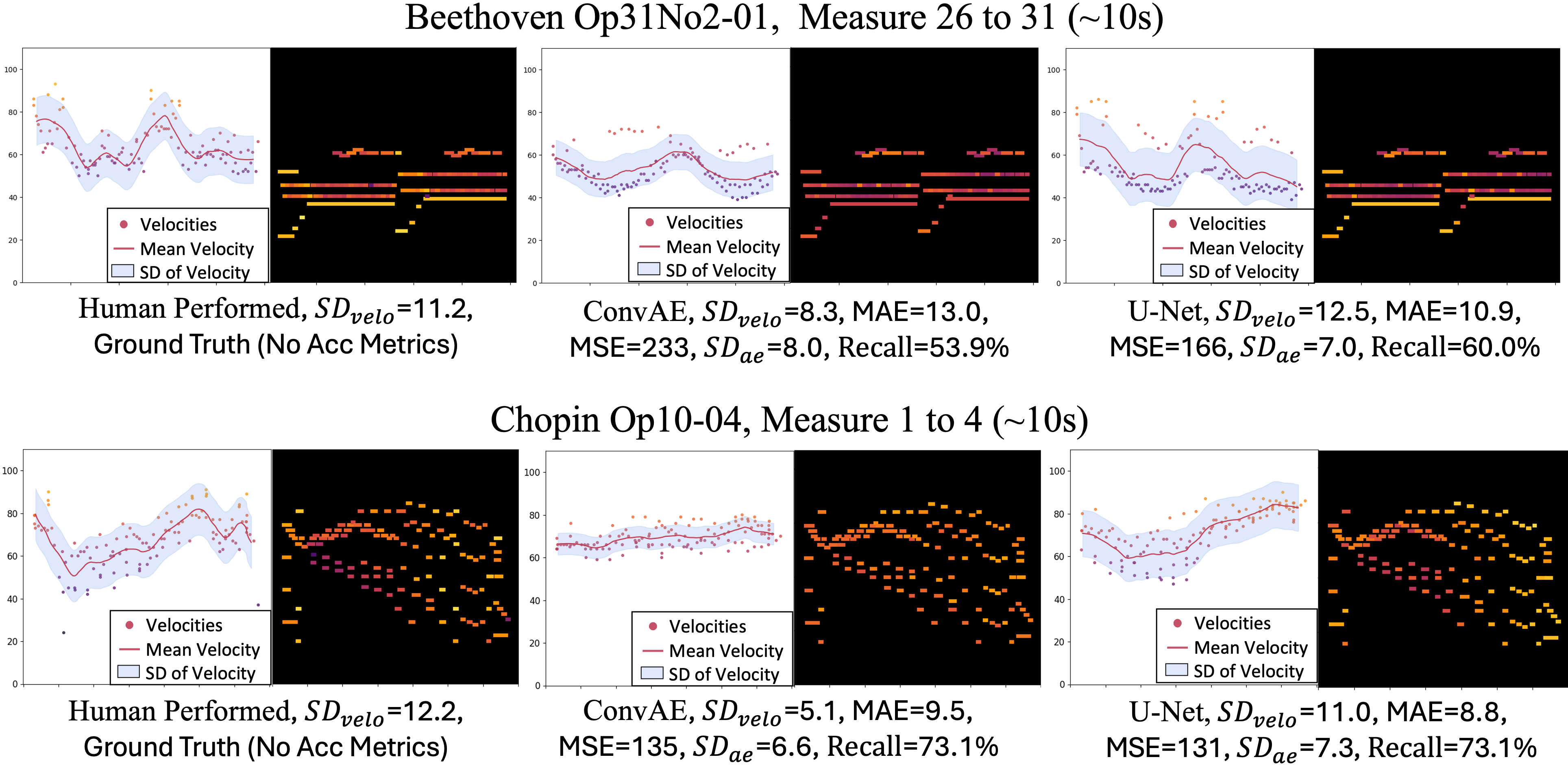}
\caption{Comparison of human-performed velocity with ConvAE and U-Net predictions. The U-Net result is more human-like than ConvAE, but for Chopin Op. 10 No. 4, accuracy metrics fail to reflect this.} 
\label{fg:visual}
\end{center}
\vspace{-24pt}
\end{figure}

\subsection{Qualitative Results through Listening Test}

A MUSHRA-like listening test was conducted to evaluate the human-likeness of performances generated by our U-Net, ConvAE \cite{kuo2021velocity}, and the Flat model.

\textbf{Participants.} We recruited 11 expert listeners (aged $\geq 18$) based on their substantial musical background and experience with critical listening studies. Participants completed survey anonymously through the Qualtrics platform \cite{qualtrics}.

\textbf{Stimuli.} The test involved eight 10s MIDI segments, each selected from a performance listed in Table 1, with human-performed velocities removed for uniform model inputs. Outputs from Flat, ConvAE, and U-Net were rendered as stimuli, while the original MIDI (with human-performed velocity) served as a reference, all using the PianoTeq 8 plug-in with the Steinway Model D instrument.

\textbf{Calibration.} The dataset used for the human-performed MIDI also contained corresponding recordings of the Yamaha Disklavier used in the performance. This is the audio that the pianist would have heard at the time of the performance. To most accurately render the MIDI, we calibrated the piano modeling software, PianoTeq 8\cite{pianoteq8}, to match those Disklavier recordings. This was achieved by adjusting the "Dynamics" control in PianoTeq 8 and comparing the Momentary Loudness via correlation coefficient and Loudness Range as described in BS.1770/EBU R128, against the reference. The results of this process presented in Table \ref{tb:pianoteq_config} indicated that a Dynamics setting of 60 dB most closely matched the audio of the original performance. This setting was used to process all files used for the test.

\vspace{-0.2cm}
\begin{table}[h]
\centering
\caption{Metrics for comparing different PianoTeq configurations to the reference.}
\begin{tabular}{|l|c|c|}
\hline
\textbf{Dynamics (dB)} & \textbf{Loudness $\Delta$ \(\downarrow\)} & \textbf{Correlation Coefficient \(\uparrow\)} \\
\hline
50  &  1.8   &    0.9583           \\ \hline
\textbf{60}  &  \textbf{0.3}   &    \textbf{0.9591} \\ \hline
70  &  0.9   &    0.9561           \\ \hline
80  &  1.9   &    0.9535           \\ \hline
90  &  3.1   &    0.9479           \\ \hline
\end{tabular}
\label{tb:pianoteq_config}
\end{table}
\vspace{-0.4cm}

\textbf{Procedure.} Participants were instructed to use headphones for accurate evaluation, as subtle velocity differences required higher playback volumes for clarity. They rated the similarity between audio rendered from model-predicted and human-performed MIDI, using a continuous slider from 0 to 5, and results were aggregated into MOS scores in Table \ref{tb:mos_scores}.

\vspace{-0.2cm}
\begin{table}[h]
\centering
\caption{Results of subjective listening test. MOS with 95\% confidence interval are reported, where "most seen" includes \{Chopin, Bach, Beethoven\}, "less seen" includes \{Rachmaninoff, Haydn, Mozart\}, and "unseen" includes \{Bartók, Ravel\}.}
\label{tb:mos_scores}
\begin{tabular}{lcccc}
\toprule
\textbf{Model} & \textbf{Most Seen MOS \(\uparrow\)} & \textbf{Less Seen MOS \(\uparrow\)} & \textbf{Unseen MOS \(\uparrow\)} & \textbf{Overall MOS \(\uparrow\)} \\
\midrule
Flat   & $1.58 \pm 0.36$ & $1.64 \pm 0.44$ & $1.18 \pm 0.42$ & $1.50 \pm 0.37$ \\
ConvAE & $1.93 \pm 0.34$ & $2.40 \pm 0.39$ & $1.80 \pm 0.44$ & $2.08 \pm 0.33$ \\
\textbf{U-Net}  & $\mathbf{3.10 \pm 0.38}$ & $\mathbf{3.16 \pm 0.29}$ & $\mathbf{2.67 \pm 0.46}$ & $\mathbf{3.01 \pm 0.34}$ \\
\bottomrule
\end{tabular}
\end{table}
\vspace{-0.2cm}

\textbf{Results.} Although a gap to human performance remains (no model achieved a perfect similarity score of 5), our U-Net model significantly outperformed other approaches. The "Flat" model performed worst, aligning with the survey in previous work \cite{kuo2021velocity}. A key limitation in this survey was distinguishing whether performance was impacted by model familiarity with composer styles (seen vs. unseen) or by the music's complexity. Despite this ambiguity, the consistently narrow 95\% confidence intervals across all groups indicate that listeners provided similar ratings, reinforcing the overall reliability of this survey.

\section{Conclusion}

In this paper, we propose a U-Net model for MIDI velocity prediction inspired by image colorization. Our model integrates window attention and a task-specific loss function to address the sparse nature of image-like MIDI representations, and we also explore the impact of MIDI segment duration. Both objective and subjective evaluations confirmed the inadequacy of default MIDI velocities and demonstrated that the proposed U-Net outperforms existing methods. Furthermore, we used a visual representation to assess the reliability of various objective metrics, highlighting \(SD_{\text{velo}}\) as a particularly effective indicator. The potential of \(SD_{\text{velo}}\) for applications beyond evaluation, such as in model training, warrants further investigation.

A key limitation is that the proposed model has only been validated on piano data. Although a cross-dataset evaluation was conducted, generalization across different composer styles and other instruments remains unexplored. Future work should investigate this area, as image-based models may offer advantages over sequential models that rely on instrument-specific patterns (i.e., linguistic information). This makes them a promising direction for further research.

\begin{credits}
\subsubsection{\ackname} This work was conducted during Zhanhong He's research internship at Dolby Australia. We would like to thank the Dolby staffs for their support with the subjective listening test, and the other intern Hanyu Meng for selecting the MIDI segments.

\subsubsection{\discintname}
The authors declare no competing interests to declare that are relevant to the content of this article.
\end{credits}

\bibliographystyle{splncs04}
\bibliography{cmmr2025}
%





\end{document}